\documentstyle[mnras_cite,epsfig,eqnarray]{mn2e}
\title[Black Hole Spin Properties of 130 AGN]{Black Hole Spin Properties of 130 AGN}
\author[R. A. Daly et al.]{Ruth.~ A.~ Daly \thanks{E-mail:
rdaly@psu.edu} 
and Trevor~B.~Sprinkle \\
Penn State University, Reading, PA 19608, USA} 

\begin{document}

%\date{}

%\pagerange{\pageref{firstpage}--\pageref{lastpage}} \pubyear{2002}

\maketitle

\label{firstpage}

\begin{abstract}
Supermassive black holes may be described by their mass and spin. 
When supermassive black holes are active, the activity provides
a probe of the state of the black hole system.  
The spin of a hole can
be estimated when the black hole mass and beam power of the source are known
for sources with powerful outflows. 
Seventy-five sources for 
which both the black hole mass and beam power could be obtained are
identified and used to obtain estimates of black hole spins. The 
75 supermassive black holes studied include 52 FRII radio galaxies and 
23 FRII radio loud quasars with redshifts ranging from about zero to 
two. The new values are combined
with those obtained previously for 19 FRII radio galaxies, 7 FRII radio loud
quasars, and 29 radio sources associated with CD galaxies to form
samples of 71 FRII radio galaxies, 30 FRII quasars, and a 
total sample of 130 spin values; 
all of the sources are associated with massive elliptical galaxies. 
The new values obtained are similar to those obtained earlier
at similar redshift, and range from about 0.1 to 1 for FRII sources.
The overall results are consistent with those obtained 
previously: the spins tend to decrease with decreasing redshift
for the FRII sources studied. There is a hint that the range of values of
black hole spin at a given redshift is larger for FRII quasars than for FRII 
radio galaxies. There is no indication of a strong correlation between
supermassive black hole mass and spin for the supermassive black holes
studied here. The relation between beam power and black hole mass is
obtained and used as a diagnostic of the outflows 
and the dependence of the magnetic field strength on black hole mass.

\end{abstract}

\begin{keywords} {black hole physics -- galaxies: active}
\end{keywords}

\section{INTRODUCTION}

The study of spins of supermassive black holes provides insight into the
merger and accretion history of supermassive black holes 
(e.g. Hughes \ Blandford 2003; Volonteri et al. 2005;  
King \& Pringle 2006; King \& Pringle 2007; 
Volonteri, Sikora, and Lasota (2007); King, Pringle, \& Hofmann (2008); 
Berti \& Volonteri 2008; Lagos et al. 2009; 
Barausse 2012; Dubois, Volonteri, \& Silk 2013).  
Significant progress has been made on estimating spins of supermassive
black holes (e.g. Reynolds 2013; Gnedin et al. 2012; Daly 2011). 
For example, when X-rays are observed 
from the inner region of the accretion disk of an AGN, the properties of the 
disk's emission spectrum may be used to estimate the black hole spin 
(e.g. Gallo et al. 2005; Brenneman \& Reynolds 2006; Crummy et al. 2006; 
Miniutti et al. 2007; Miniutti et al. 2009; Schmoll et al. 2009; 
Zoghbi et al. 2010; Pérez 2010; Gallo et al. 2011;
Patrick et al. 2011; Brenneman et al. 2011; Patrick et al. 2012; 
Walton et al. 2012; Tatum et al. 2012)  
using the method proposed by Fabian et al. (1989). 
When an AGN has a powerful outflow, the properties of the outflow 
may be used 
to estimate the spin of the AGN (Daly 2009a,b, 2011; Gnedin et al. 2012). 
Spins of stellar mass black holes can be studied using the 
continuum-fitting  method (Zhang et al. 1997) as summarized by 
McClintock, Narayan, \& Steiner (2013). 
Of course, black holes associated with sources that 
are not X-ray emitters and holes associated with sources without outflows 
may have substantial spins, but sources that are X-ray emitters or
have outflows provide an opportunity to empirically estimate the 
spin of the black hole.  

Black hole spin values that have been reported 
range from about 0.2 to 1 for determinations using outflows 
from 26 powerful FRII (Fanaroff \& Riley 1974) sources (Daly 2011);  
from about 0.4 to just less than one using outflows from 14 nearby AGN 
(Gnedin et al. 2012); from about 0.01 to about 0.4 for outflows from
supermassive black holes associated with CD galaxies (Daly 2011);  
from about zero to greater than 0.95 
using continuum-fitting for 10 stellar mass black holes
(Green et al. 2001; Orosz 2003; Davis et al. 2006; McClintock et al. 2006; 
Shafee et al. 2006; Orosz et al. 2007; Liu et al. 2008; Gou et al. 2009;
Orosz et al 2009; Cantrell et al. 2010; Gou et al. 2010; 
Gou et al. 2011; Orosz et al. 2011a; Orosz et al. 2011b; 
Steiner et al. 2011; Steiner et al. 2012;
Steeghs et al. 2013);   
and from about 0.5 to just less than 1 for 22 
AGN with detailed X-ray spectra of the accretion disk  
(as summarized by Reynolds 2013), though it should be noted that
additional AGN studied using the same method have low (undetected) spin values 
and are consistent with little or no spin (Patrick et al. 2012).   

Here, the method proposed by Daly (2009b, 2011) 
is applied to 75 new sources. The method is reviewed in 
section 2, the results are described in section 3, and 
are summarized and discussed in section 4. 

A standard Lambda Cold Dark Matter cosmology with zero redshift 
mean mass density relative to the critical value of  
$\Omega_{\rm m} = 0.3$, normalized cosmological constant of 
$\Omega_{\Lambda} = 0.7$, zero space curvature, 
and a value of Hubble's constant of 
70 km/s/Mpc is assumed throughout. 

\section{The Method}

The method used here to estimate spins of black holes is valid when the
outflow is powered, at least in part, by the spin of the supermassive
black hole.  A class of models have been proposed and developed in which 
the jets are powered in part or in full by the spin energy associated with 
a rotating black hole and surrounding region (e.g. Blandford \& Znajek 1977; 
Macdonald \& Thorne 1982; Rees 1984; Begelman, Blandford, \& Rees 1984; 
Thorne et al. 1986; Punsly \& Coroniti 1990; Blandford 1990; 
Meier 1999, 2001; Koide et al. 2000; McKinney \& Gammie 2004; 
De Villiers et al. 2005; 
Hawley \& Krolik 2006; Reynolds, Garofalo, \& Begelman 2006; 
Tchekhovskoy, Narayan, \& McKinney 2011; 
Penna et al. 2010; Narayan, McClintock, \& Tchekhovskoy 2013). 
In many of these models the relationship between black hole spin,  
beam power of the 
outflow, the braking magnetic field strength, and black 
hole mass is identical expect for a constant
of proportionality. They satisfy 
\begin{equation}
L_j \propto j^2M^2B^2 
\end{equation}
(e.g. Blandford \& Znajek 1977; 
Blandford 1990; Meier 1999; Reynolds et al. 2006) 
where $L_j$ is the beam power of the 
outflow, $M$ is the black hole mass, 
$j$ is the black hole spin, $j=a/m$ where $a=S/(Mc)$, 
$m = GM/c^2$, $S$ is the spin angular momentum, 
$c$ is the speed of light, and $B$ is the 
the strength of the poloidal component of the magnetic
field threading the accretion disk and black hole ergosphere. 
This relationship implies that the black hole spin $j$ may be determined
where the beam power, black hole mass, and braking magnetic field strength
are known or can be estimated (Daly 2009b, 2011):  
\begin{equation}
j = \kappa (L_{44})^{0.5} ~B_4^{-1} ~M_8^{-1}
\end{equation}
where $L_{44}$ is the beam power in units of $10^{44}$ erg/s, $B_4$ is the poloidal component of 
the magnetic field in units of $10^4$ G, and $M_8$ 
is the black hole mass in units of $10^8 \rm{M}_{\odot}$. The constant of proportionality $\kappa$ 
varies by a factor of a few for different models; for example, in the hybrid model of Meier (1999) 
$\kappa \approx (1.05)^{-1/2}$, while in the model of Blandford \& Znajek (1977) $\kappa \approx \sqrt{5}$.The results shown here are obtained in the hybrid model of Meier (1999), and can easily
be scaled to other models.

Recent work on stellar mass black holes provides the first direct empirical evidence
that eq. (1) provides a good description of the relationship between black hole
spin and beam power.  Narayan \& McClintock (2012) studied five stellar mass black holes 
with independently determined
spin and outflow power and showed that the power was proportional to the square of
the spin as described by eq. (1). 

Black hole spins are estimated here for sources with empirical determinations of 
beam power and black hole mass.  
To obtain these estimates, the same three characterizations of the 
magnetic field strength studied by Daly (2011) are considered; these include
an Eddington magnetic field 
strength, a constant magnetic field strength, and a magnetic field 
strength that is proportional to the black hole spin. These three
field strengths are related to the black hole properties in 
different ways.

\section{Results}
\subsection{The Samples}

Samples of sources for which both the beam power $L_j$ and black hole
mass $M$ can be empirically determined are selected for study. Beam powers
can be determined for powerful classical double radio sources 
using multifrequency radio images of the radio lobe regions of
the sources (e.g. Alexander \& Leahy 1987; Leahy, Muxlow, \& Stephens 1989;
Liu, Pooley, \& Riley 1992; Wellman, Daly, \& Wan 1997; Guerra, Daly,
\& Wan 1998; O'Dea et al. 2009). These AGN 
are known as FRII sources (Fanaroff \& Riley 1974) and are 
hosted by massive elliptical galaxies (e.g. 
Lilly \& Longair 1984; Best, Longair \& R\"ottgering
1998; Bettoni et al. 2001; Dunlop et al. 2003; McLure et al. 2004).
The radio sources are edge-brightened with most of the radio power of the source
coming from hot spots at the outer regions of the cigar-shaped radio emitting region.
The sources are large, often 
much larger than the opitcal size of the host galaxy (e.g. Leahy, Muxlow, \&
Stephens 1989). The sources are producing synchrotron 
radiation isotropically, so there are no orientation-dependent 
corrections for relativistic 
beaming and boosting of the emission due to bulk motion of the emitting material.  
The beam power carried away from the AGN
is dumped at the extremities of the source at the radio hot spots. The radio structure
of the sources indicates that the sources are growing supersonically
(e.g. Leahy, Muxlow, \& Stephens 1989), so the equations of strong shock 
physics can be applied to the sources. This allows the beam power to be determined
very cleanly by applying the equations of strong shock physics to the outer region
of the source using parameters  
determined using multifrequency radio data of the 
extended radio emitting region (e.g. Alexander \& Leahy 1987; Leahy, Muxlow, \& Stephens 1989;
Liu, Pooley, \& Riley 1992; Wellman, Daly, \& Wan 1997; Guerra, Daly,
\& Wan 1998; O'Dea et al. 2009). In this application,
the one parameter that typically must be assumed to describe the plasma, 
the deviation from minimum energy conditions, fortuitously cancels so the beam power 
may be obtained in a model independent manner 
(O'Dea et al. 2009). 

Daly (2011) studied samples of 19 FRII 
radio galaxies (RG) and 7 FRII radio loud quasars (RLQ) which had
beam powers determined using multifrequency radio images of the 
extended radio lobes and known black hole masses and determined and
studied spin values for these sources. To extend this sample,
Daly et al. (2012) used a sample of 31 powerful FRII sources to determine
the relationship between beam power and radio power that is applicable for 
FRII sources; all of these sources had beam powers determined with 
multifrequency radio data as described above. The relationship
between beam power and radio power was found to be independent of how 
radio power was defined and of whether certain sources were included or excluded,
and can be summarized as ${\rm Log}(L_{44}) = (0.83 \pm 0.14) {\rm Log}(P_{44}) + (2.09 \pm 0.06)$,
where $L_{44}$ is the beam power in units of $10^{44} {\rm erg/s}$ and $P_{44}$ is the 
product of the rest frame radio power at 178 MHz and 178 MHz in units
of $10^{44} {\rm erg/s}$. 

The rather large uncertainty of the slope of the relation
reflects the fact that there is substantial scatter in the relationship. 
Indeed, this is borne out by the detailed numerical modeling and analysis of FRII sources 
by Hardcastle \& Krause (2013). Hardcastle \& Krausse (2013) 
find a substantial amount
of scatter in the beam power - radio luminosity relation in their simulated sources and 
conclude that this is likely due to the dependence of
source properties on the environment, as is known from direct observations of
sources (e.g. Barthel \& Arnaud 1996; Belsole et al. 2007). 
The sources studied by Daly et al. (2012) had redshifts 
fairly evenly distributed between zero and two. The ratio of beam power to radio power 
and scatter in the relationship was
found to be independent of redshift, indicating that a range of environments at a
given redshift rather than systematic evolution of environments with 
redshift was the cause of the large scatter in the derived relationship.

FRII sources with known
supermassive black hole mass and radio power but for which multifrequency maps
were not avaible to determine the beam power directly were identified. 
Beam powers for these sources were obtained by applying the relationship 
between beam power and radio power applicable to FRII sources, 
listed above. This led to 52 radio galaxies 
and 23 radio loud quasars for which new black hole spins could be estimated; these 
sources and their properties are listed in Tables 1 and 2. The uncertainties of these
beam powers are 
obtained by propagating the uncertainties in the derived relationship
between beam power and radio power. 

In addition to the FRII sources studied earlier, 29 extended radio sources
from Rafferty et al. (2006), which are primarily 
FRI sources, were also studied by Daly (2011). 
Beam powers for these sources were obtained by Rafferty et al. (2006) by 
combining the radio 
source properties with the properties of the source environments. 
Thus, the total sample to be 
studied here includes 71 FRII RG, 30 FRII RLQ, and 29 radio sources associated 
with CD galaxies. 

The beam powers and black hole masses for the FRII RG are shown in Fig. 1; those
for FRII RLQ are shown in Fig. 2; and those for 29 sources associated with CD galaxies
are shown in Fig. 3. The properties of the previously studied sources are listed
in Tables 1, 2, and 3 of Daly (2011). All of the sources are associated with massive 
elliptical galaxy hosts.

\begin{figure}
    \centering
    \includegraphics[width=80mm]{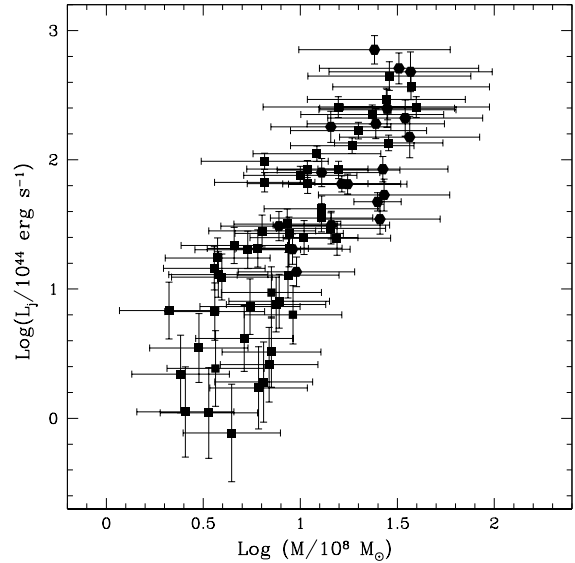}

\caption{Beam powers and black hole masses for the 52 new FRII 
RG (solid squares) 
and 19 previously studied FRII RG (solid circles) are shown 
here. }  

	  \label{fig:F1}
    \end{figure}

\begin{figure}
    \centering
    \includegraphics[width=80mm]{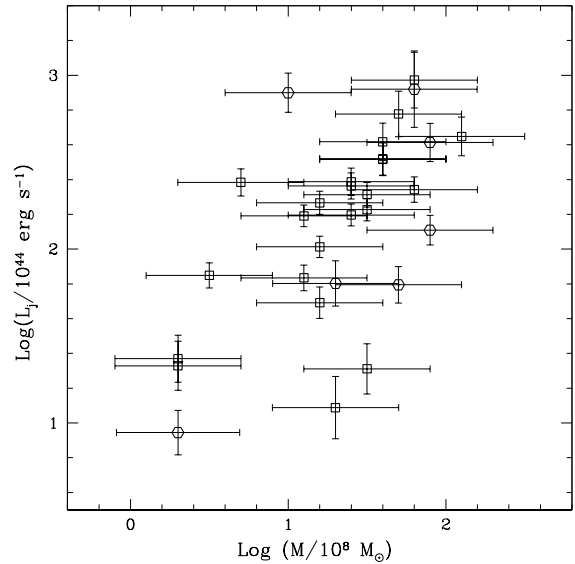}

\caption{Beam powers and black hole masses for 
23 new FRII RLQ (open squares) and 7 previously studied 
FRII RLQ (open circles) are shown here.} 
	  \label{fig:F2}
    \end{figure}

\begin{figure}
    \centering
    \includegraphics[width=80mm]{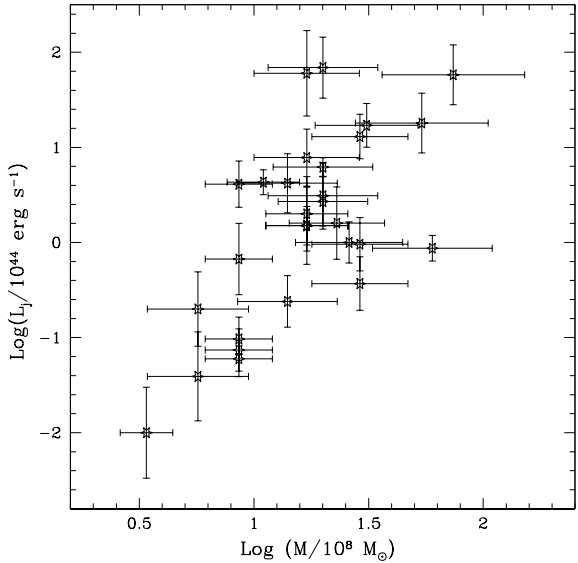}

\caption{Beam powers and black hole masses for  
29 radio sources associated with CD galaxies from Rafferty et al. (2006), 
previously studied by Daly (2011), are shown here.} 
	  \label{fig:F3}
    \end{figure}

\subsection{Determination of Black Hole Spin}
The black hole spins are obtained using eq. (2) given the beam powers
and black hole masses listed in Tables 1 and 2. 
Three magnetic fields are 
considered as described by Daly (2011) and include 
a constant field strength, $B_4 = 1$, an Eddington 
magnetic field strength, $B_{4,EDD} \simeq 6 M_8^{-1/2}$,
and a field strength proportional to black hole spin $B_4 \simeq 2.78 j$. 
The choice of $B_4=1$ is motivated by the empirical results of 
Piotrovich et al. (2011) who find that this is a typical value of
accretion disk field strengths and by theoretical considerations 
(e.g. Blandford 1990); the choice of an Eddington magnetic field
strength is motivated by theoretical and empirical considerations
(e.g. King 2010); and the choice of $B \propto j$ is motivated
by empirical results obtained with powerful FRII radio sources
(Daly \& Guerra 2002; Daly et al. 2009).   
The resulting spins and
their uncertainties are listed in Tables 1 and 2; the uncertainties
are obtained in the standard manner, 
as described in section 3.2 of Daly (2011). 

The black hole spins obtained here are combined with those obtained previously 
(Daly 2011) yielding samples of 71 FRII radio galaxies and 30 FRII radio 
loud quasars. The spins are shown as a function of $(1+z)$  
in Figs. 4 - 9 for the three magnetic field strengths considered. 
Best fit lines are obtained for the FRII sources and the slopes
of these lines and best fit parameters for the combined sample of 
galaxies and quasars are listed in Table 3. Note that the slope is independent
of the value of $\kappa$ in equation 1, and thus is independent of the specific
model of spin energy extraction. 

The new results obtained here are compared with those obtained previously with 
19 FRII radio galaxies and 7 FRII radio loud quasars in columns (2) and (3)
of Table 4. The slopes obtained with the full sample studied here are within one sigma
of those obtained previously.

\begin{figure}
    \centering
    \includegraphics[width=80mm]{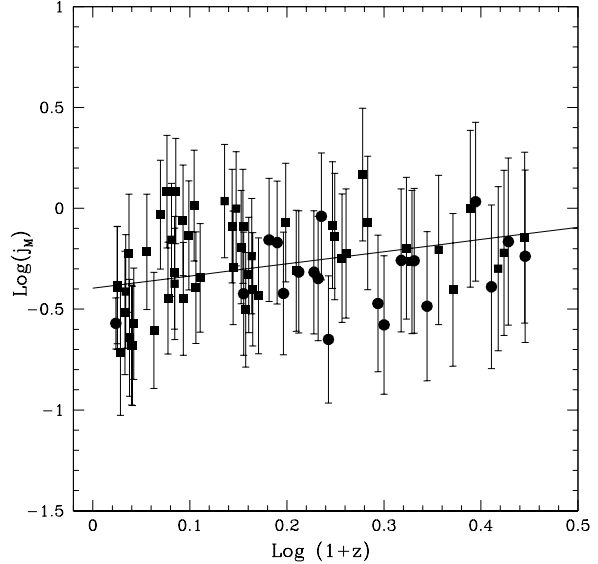}

\caption{Black hole spins obtained with a constant magnetic 
field strength of $10^4$ G for the 52   
FRII radio galaxies listed in Table 1 (shown as solid squares) 
and the  19 previously studied FRII radio galaxies
(shown as solid circles). The best fit line indicates that
${\rm Log} (j) = (0.60 \pm 0.22)~{\rm Log}(1+z)+ (-0.40 \pm 0.04)$.}
		  \label{fig:F4}
    \end{figure}

\begin{figure}
    \centering
    \includegraphics[width=80mm]{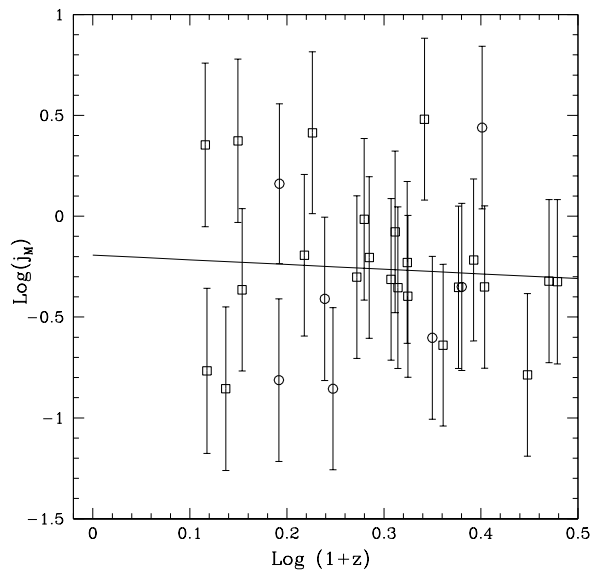}

\caption{Black hole spins obtained with a constant magnetic 
field strength of $10^4$ G  for the 23   
FRII radio loud quasars listed in Table 2 (shown as open squares) 
and the 7 previously studied FRII radio loud quasars 
(shown as open circles).
 The best fit line indicates that
${\rm Log} (j) = (-0.23 \pm 0.71)~{\rm Log}(1+z)+ (-0.19 \pm 0.22)$.}
		  \label{fig:F5}
    \end{figure}

\begin{figure}
    \centering
    \includegraphics[width=80mm]{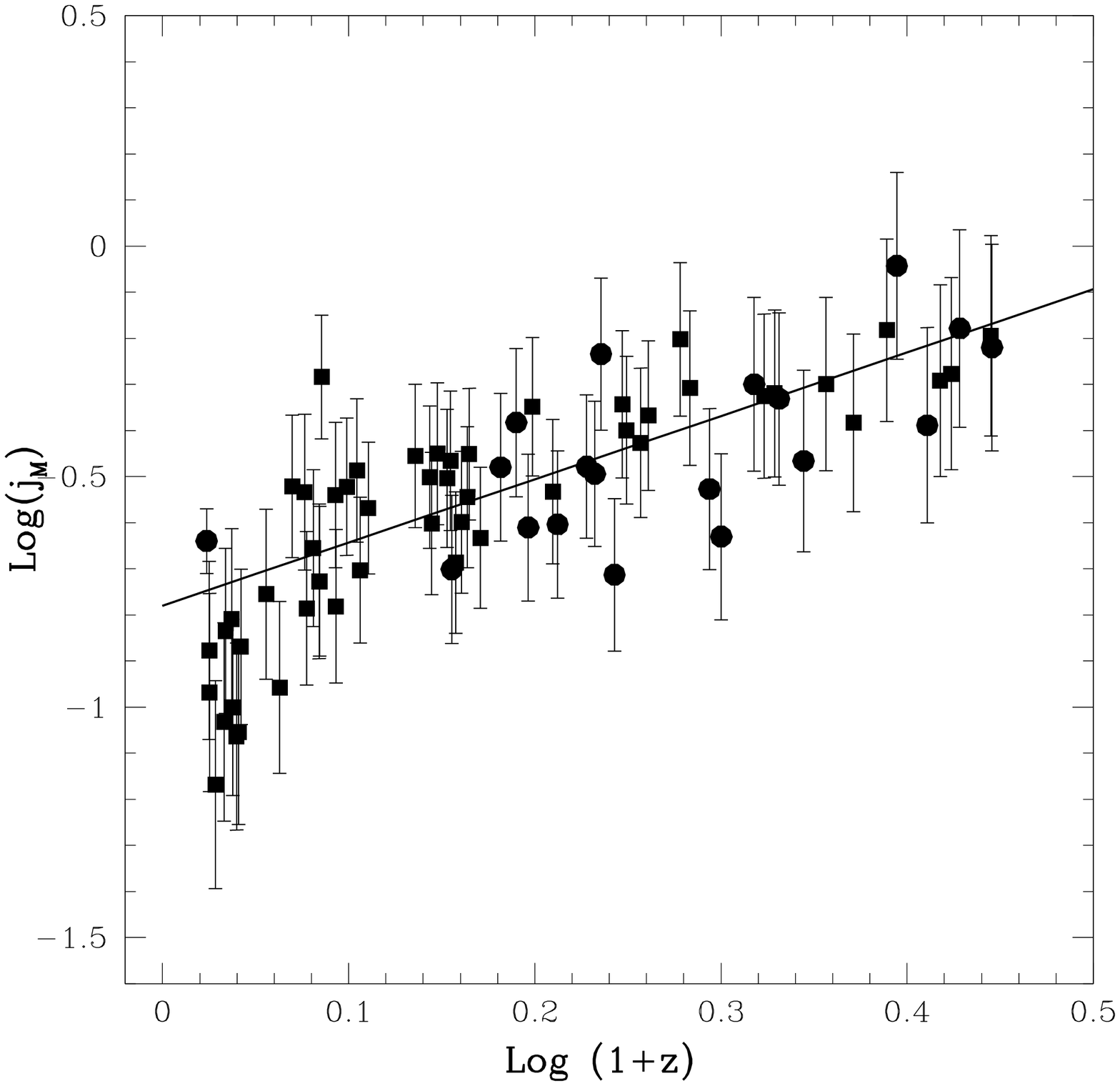}
	 
\caption{Black hole spins obtained with an Eddington magnetic field strength
for the sources shown in Fig. \ref{fig:F4}; the symbols 
are the same as in that Figure and the best fit line indicats that
${\rm Log} (j) = (1.37 \pm 0.15)~{\rm Log}(1+z)+ (-0.78 \pm 0.03)$.}
	  
\label{fig:F6}
    \end{figure}

\begin{figure}
    \centering
    \includegraphics[width=80mm]{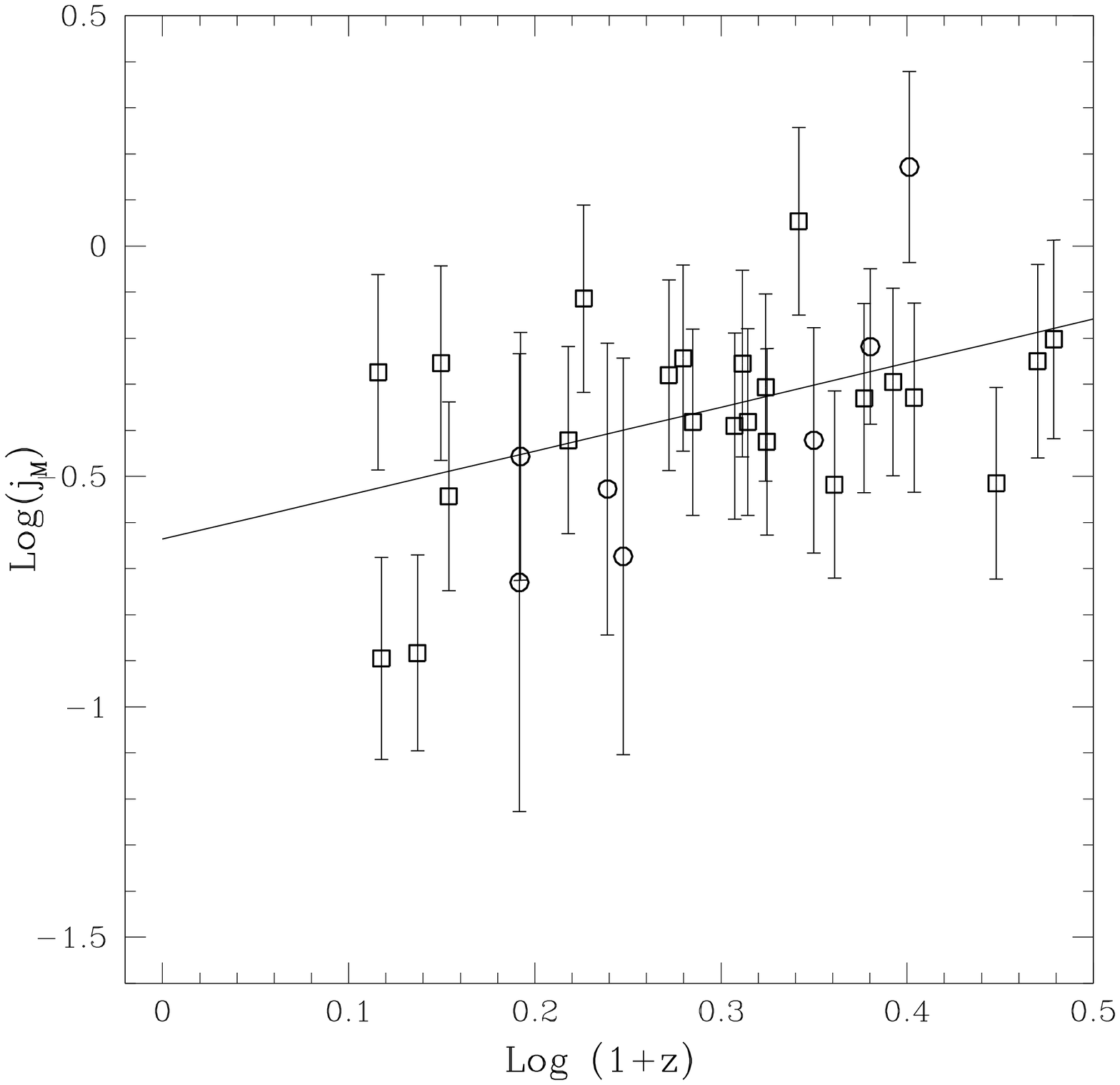}

\caption{Black hole spins obtained with an Eddington magnetic field strength
for the sources shown in Fig. \ref{fig:F5}; the symbols 
are the same as in that Figure and the best fit line indicates that 
${\rm Log} (j) = (0.96 \pm 0.36)~{\rm Log}(1+z)+ (-0.64 \pm 0.11)$.}
	  
\label{fig:F7}
    \end{figure}

\begin{figure}
    \centering
    \includegraphics[width=80mm]{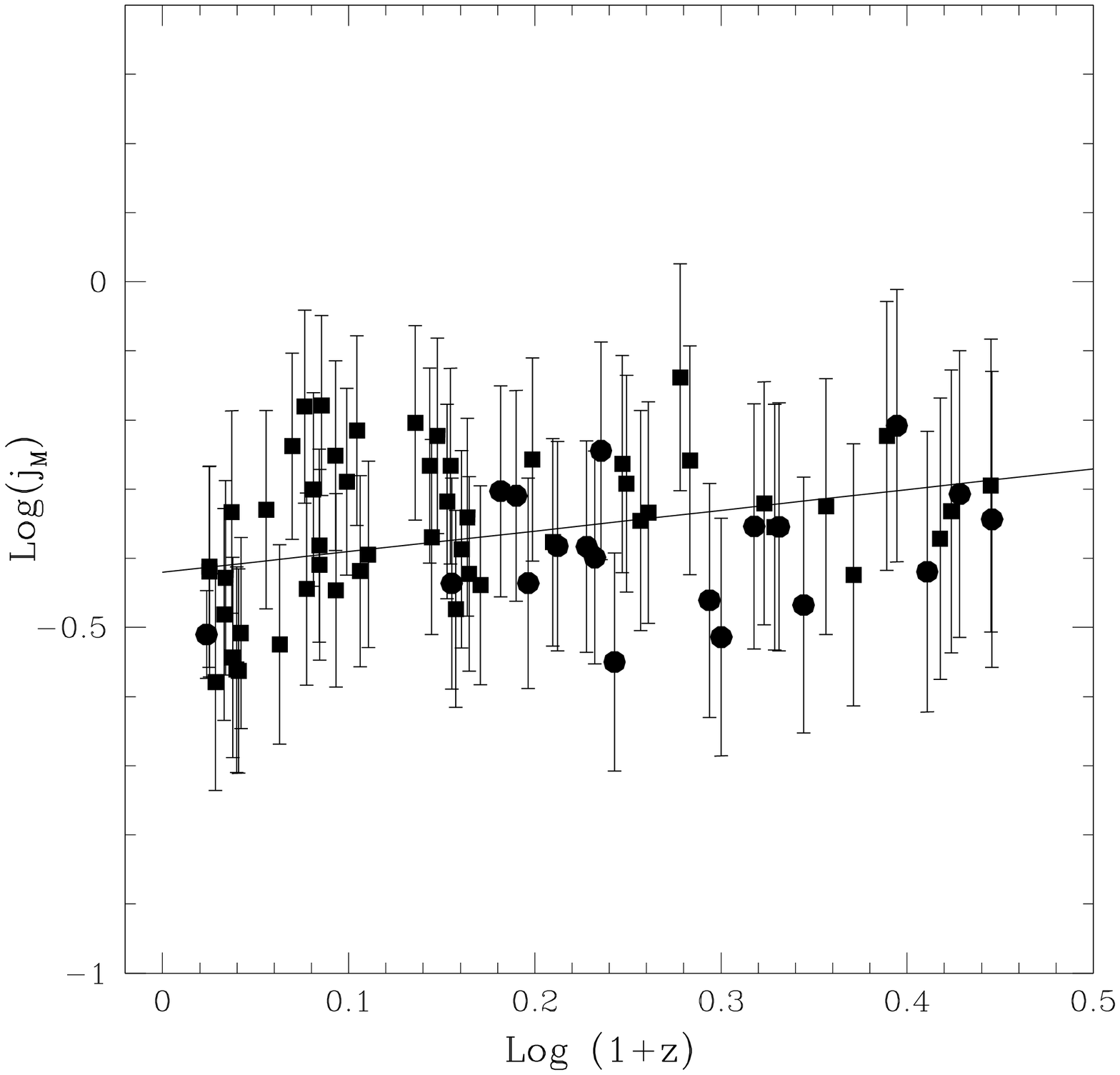}
			 
\caption{Black hole spin obtained with a magnetic field 
strength that is proportional to the spin
for the sources shown in Fig. \ref{fig:F4}; the symbols 
are the same as in that Figure and the best fit line indicates that 
${\rm Log} (j) = (0.30 \pm 0.11)~{\rm Log}(1+z)+ (-0.42 \pm 0.02)$.}
		  \label{fig:F8}
    \end{figure}

\begin{figure}
    \centering
    \includegraphics[width=80mm]{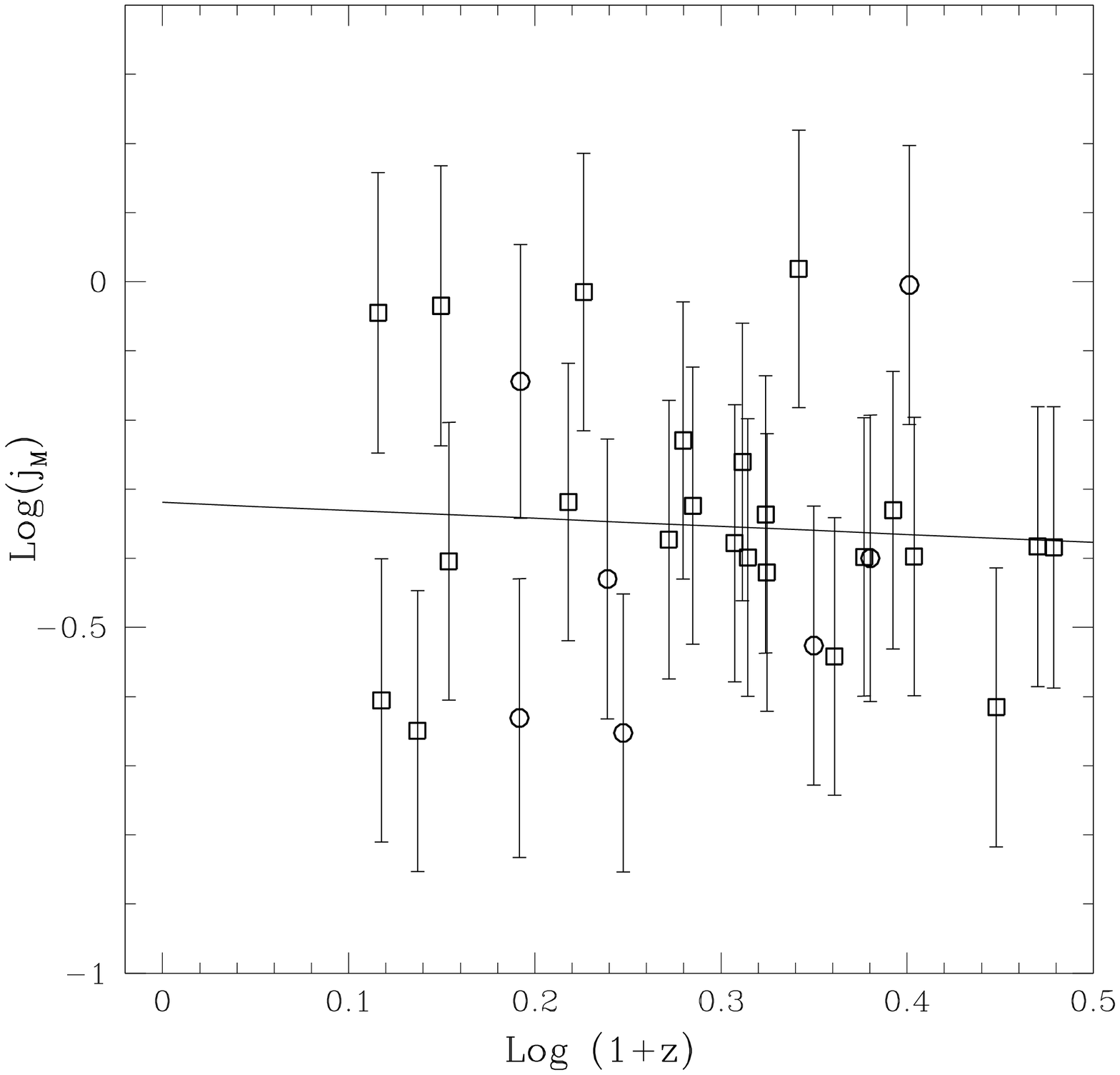}
			 
\caption{Black hole spin obtained with a magnetic field 
strength that is proportional to the spin for the sources
shown in Fig. \ref{fig:F5}; the symbols 
are the same as in that Figure and the best fit line indicates that
${\rm Log} (j) = (-0.12 \pm 0.36)~{\rm Log}(1+z)+ (-0.32 \pm 0.11)$.}
		  \label{fig:F9}
    \end{figure}

\begin{figure}
    \centering
    \includegraphics[width=80mm]{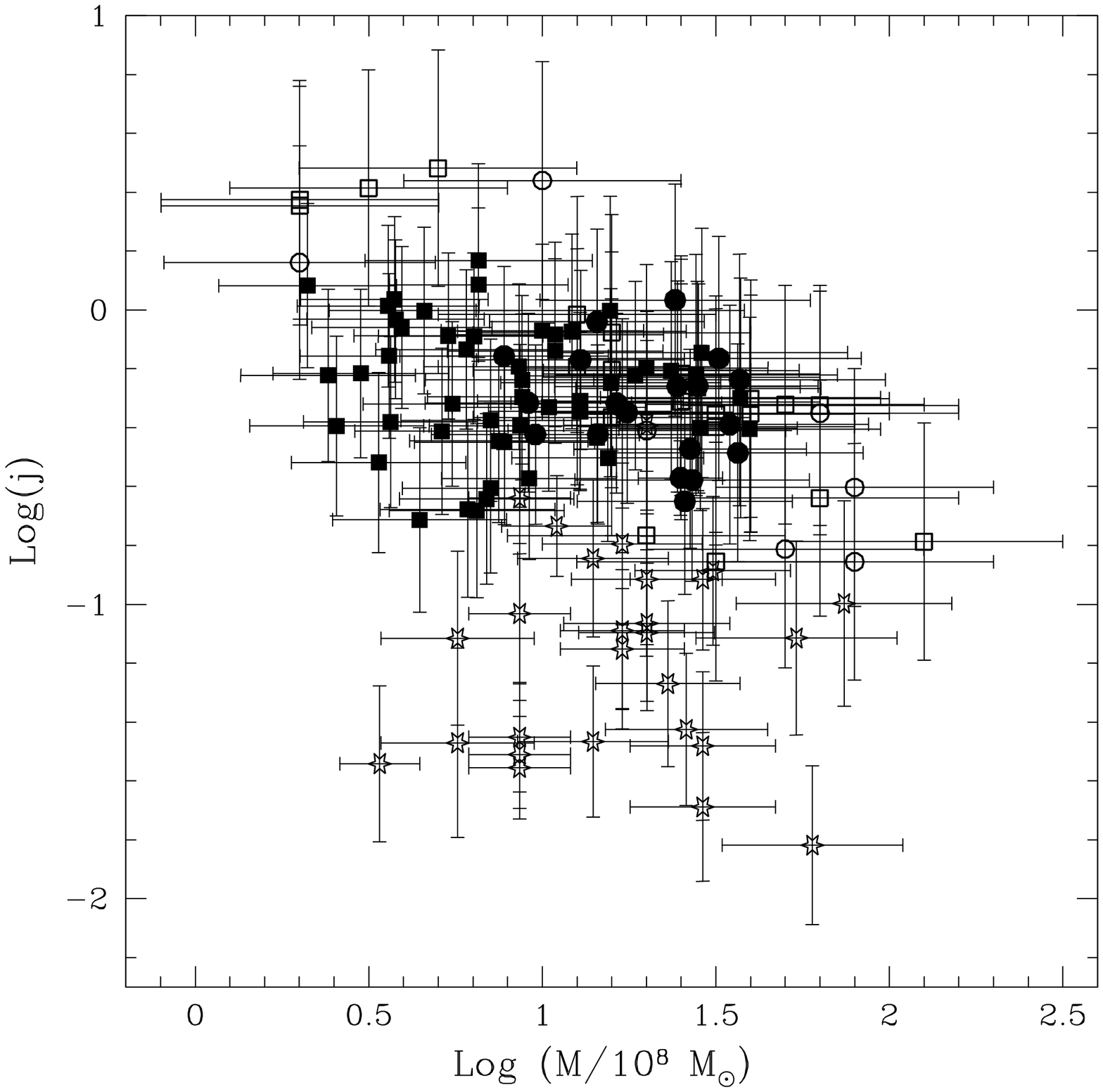}
			 
\caption{Black hole spin obtained with a constant magnetic 
field strength of $10^4$ G is shown as a function of black hole mass; 
the symbols are as in Figs. \ref{fig:F1}, \ref{fig:F2}, and \ref{fig:F3}. 
An unweighted best line 
fit to the 101 FRII radio galaxies and quasars yields a slope of 
$-0.32 \pm 0.06$, while a weighted fit including the uncertainty of 
spin but not that of mass yields a slope of $-0.30 \pm 0.06$ with 
a $\chi^2$ of 51 for 99 degrees of freedom. }
		  \label{fig:F10}
    \end{figure}

\begin{figure}
    \centering
    \includegraphics[width=80mm]{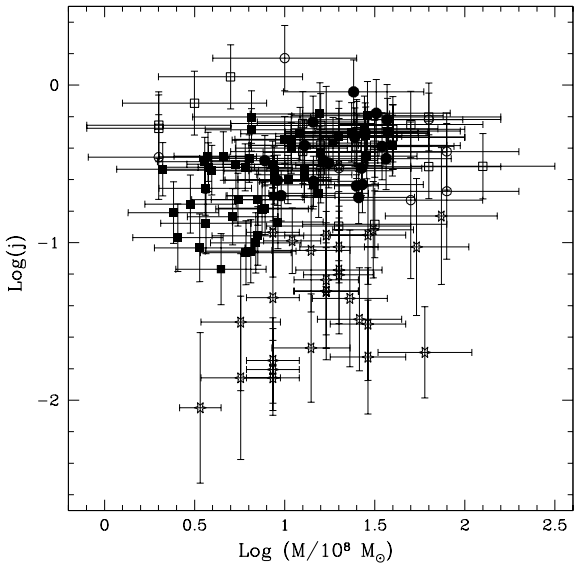}
			 
\caption{Black hole spins obtained with an Eddington magnetic field strength
is shown as a function of black hole mass; 
the symbols are as in Figs. \ref{fig:F1}, \ref{fig:F2}, and \ref{fig:F3}.
An unweighted best line 
fit to the 101 FRII radio galaxies and quasars yields a slope of 
$+0.19 \pm 0.06$, while a weighted fit including the uncertainty of 
spin but not that of mass yields a slope of $+0.18 \pm 0.06$ with 
a $\chi^2$ of 155 for 99 degrees of freedom.  }
		  \label{fig:F11}
    \end{figure}

\begin{figure}
    \centering
    \includegraphics[width=80mm]{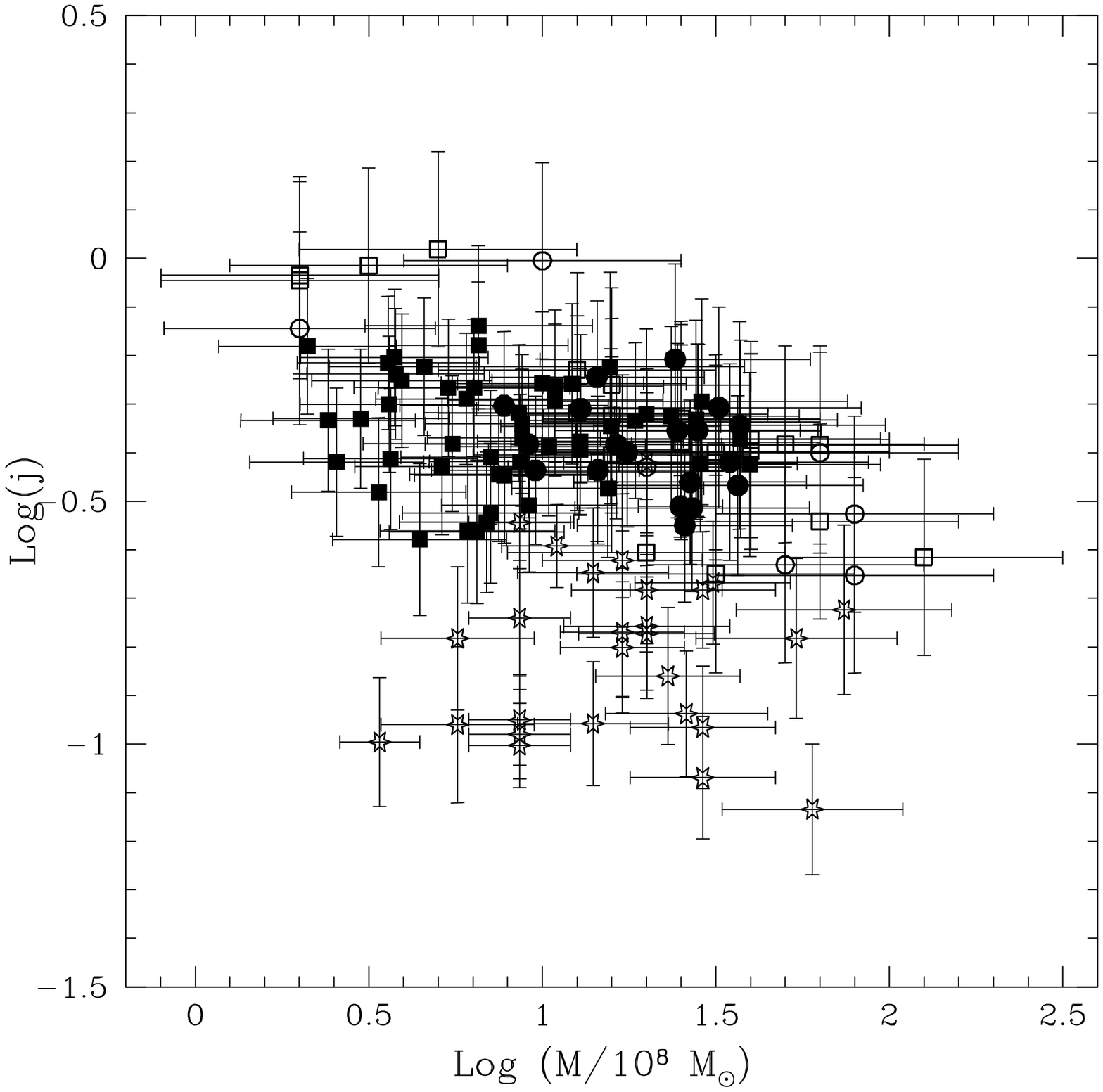}
			 
\caption{Black hole spin obtained with a magnetic field 
strength that is proportional to the spin is 
shown as a function of black hole mass; 
the symbols are as in Figs. \ref{fig:F1}, \ref{fig:F2}, and \ref{fig:F3}.
An unweighted best line 
fit to the 101 FRII radio galaxies and quasars yields a slope of 
$-0.16 \pm 0.03$, while a weighted fit including the uncertainty of 
spin but not that of mass yields a slope of $-0.15 \pm 0.03$ with 
a $\chi^2$ of 51 for 99 degrees of freedom. }
		  \label{fig:F12}
    \end{figure}

\subsection{The Beam Power - Black Hole Mass Connection}

Equation (1) indicates that the beam power is expected to be proportional
to the square of the 
black hole mass unless the magnetic field strength is related to
the black hole mass when the outflow is powered at least in part
by the black hole spin. The normalization of the empirical relationship
between beam power and black hole mass may provide some hints as to 
whether the outflow is powered by the spin of the black hole or by
processes associated with the accretion disk. For example, Meier (2001)
and Livio, Ogilvie, \& Pringle (1999) show that outflows from thin
accretion disks associated with black holes with masses of about
$10^9 M_{\odot}$ can produce outflows with beam powers of up to about
$10^{43}$ erg/s and including energy extracted from a spinning black hole
does not increase the beam power significantly. However, the extraction of
spin energy from a system with a geometrically thick accretion disk
can significantly increase the beam power yielding beam powers of 
$10^{45}$ erg/s for black hole masses of about $10^9 M_{\odot}$. 
In Blandford-Payne type models (Blandford \&
Payne 1982) and models developed along these lines (e.g. Livio, Pringle, 
\& King 2003), the beam power 
is expected to go roughly as the mass to the power 1.5 to 2 for
reasonable assumptions. 

An unweighted fit to the FRII RG shown in Fig. 1 indicates that 
$L_{44} = 10^{(-0.33 \pm 0.16)}M_8^{(1.79 \pm 0.14)}$, or $L_{44} \simeq 
20 ~M_9^{1.8}$ while a weighted fit
that includes the uncertainty of the beam power but not the black hole
mass yields $L_{44} = 10^{(-0.32 \pm 0.18)}M_8^{(1.40 \pm 0.15)}$, or 
$L_{44} \simeq 10 M_9^{1.4}$, where $L_{44}$ is the beam power in
units of $10^{44}$ erg/s, $M_8$ is the black hole mass in units
of  $10^8 M_{\odot}$, and $M_9$ is the black hole mass in units of 
$10^9 M_{\odot}$. Thus, the  normalizations of these fits are
consistent with the outflow being powered by spin energy extraction
from the black hole, and the exponent of the L-M relation 
is roughly consistent with a
magnetic field strength that does not depend strongly on the black hole
mass.

An unweighted fit to the FRII RLQ shown in Fig. 2 indicates that 
$L_{44} = 10^{(1.25 \pm 0.23)}M_8^{(0.67 \pm 0.17)}$, or $L_{44} \simeq 
80 ~M_9^{0.7}$ while a weighted fit
that includes the uncertainty of the beam power but not the black hole
mass yields $L_{44} = 10^{(1.53 \pm 0.20)}M_8^{(0.50 \pm 0.15)}$, or 
$L_{44} \simeq 100 M_9^{0.5}$. If the magnetic field strength is 
given by the Eddington field strength, $B \propto M^{-0.5}$, then 
we expect that $L_j \propto M$ for spin energy extraction models. 
The exponent of the L-M relation obtained suggests a field strength
that decreases with increases black hole mass, and is roughly consistent
with an Eddington field strength. Thus, the 
normalizations are consistent with the outflow
being powered by spin energy extraction from the black hole,
while the exponent of the L-M relation 
suggests a field strength that is related
to the Eddington field strength.

An unweighted fit to the radio sources associated with CD galaxies 
shown in Fig. 3 indicates that 
$L_{44} = 10^{(-2.39 \pm 0.57)}M_8^{(2.11 \pm 0.46)}$, or $L_{44} \simeq 
0.5 ~M_9^{2.1}$ while a weighted fit
that includes the uncertainty of the beam power but not the black hole
mass yields $L_{44} = 10^{(-1.28 \pm 0.65)}M_8^{(1.22 \pm 0.50)}$, or 
$L_{44} \simeq 1 M_9^{1.2}$. The results of the weighted fit are very
strongly affected by one source, A2029, and when this source is
removed the weighted fit becomes  
$L_{44} = 10^{(-2.52 \pm 0.65)}M_8^{(2.33 \pm 0.53)}$, or 
$L_{44} \simeq 0.6 M_9^{2.3}$. The normalization is quite low, 
perhaps suggesting that the outflows are powered by the 
accretion disk or a combination of disk plus spin energy
extraction.  The calculated black hole spins for these 
systems are typically quite low, generally between about 0.01 and 0.1, 
so there's not much spin energy to power the outflow in any case. 
The exponent of the L-M relation is around 2 for both the unweighted
fit and the weighted fit with one outlier removed suggesting that
the magnetic field strength is independent of the black hole mass. 

\subsection{Impact of Mass Range on Results}

In order to study whether the spin as a function of redshift is strongly 
affected by the range of black hole masses included in the study, subsets of 
the samples with a black hole masses within a limited range are considered. 
Only the 55 sources from Daly (2011) are included in this part of the study -
these sources have beam powers determined directly from detailed multifrequency
radio data. 

Subsets of the sources are identified such that only 
sources with a black hole mass within one sigma of the 
mean black hole mass of the sample of 55 sources are considered. 
The mean black hole masss of the sample is determined in three ways:
the weighted mean value of the log of the mass of the sample is obtained
and used to define Log $M_1$ and its uncertainty; 
the weighted mean value of the mass of the 
sample is used to define $M_2$ and its uncertainty; and the 
unweighted mean black hole mass of the sample is used to define $M_3$. 
The results are shown in Table 4; the 
number of FRII sources that remain in each subsample is listed along 
with the best fit slopes. Generally, the slope obtained for the 
limited mass range is consistent with that obtained with the full
sample studied; the only exception is in the case of the sample
defined by $M_3$ for an Eddington magnetic field strength, and 
the slopes only differ by slightly more than one sigma. 

Thus, black hole spin as a function of redshift determined for the FRII
sources described here does not appear to be strongly affected by the 
mass range of the black holes considered. This may also be illustrated
by considering whether the spin and mass are strongly correlated.
The black hole spin is shown as a function of black hole mass in Figs. 
10. 11 and 12 for the three magnetic fields considered. There is no indication 
of a strong correlation between spin and mass for the black holes studied 
here. Results of fits to spin as a function of mass are listed in Table 5 for
subsets of the data.

\section{Summary and Conclusion}

Black hole spins are estimated for 75 new sources 
including 52 FRII radio galaxies and 23 FRII radio 
loud quasars using the method proposed by Daly (2011) 
in the context of spin energy extraction models.
The values listed in Tables 1 and 2 are obtained with the 
value of $\kappa$ appropriate for the hybrid model of Meier (1999).
The spins obtained range from about 0.1 to 0.9 or so. 
For different values of $\kappa$, 
the precise values of the spins will shift but the redshift 
evolution of the spins and the range or scatter of the spins at 
a given redshift will remain the same, and thus these quantities do not depend
upon the specific model considered. The values obtained here are 
consistent with the model-independent lower bounds on spins of 0.1 to 0.2 
typical of powerful FRII sources (Daly 2009a). 

The new samples 
introduced here do not include any sources from the previously
studied FRII samples. The new and previously studied FRII samples are combined to form samples
of 71 FRII radio galaxies and 30 FRII radio loud quasars with black hole
spin estimates. The radio galaxy and quasar samples are large enough to
study separately, and the evolution of black hole spin with redshift
is considered for radio galaxies, radio loud quasars, and the combined sample.
The slopes obtained for radio galaxies, radio loud
quasars, and the combined sample are found to be consistent at better than one sigma. 
The slope indicates that the spins of supermassive black
holes associated with powerful FRII sources 
are decreasing with decreasing redshift for all three magnetic
field strengths considered. This trend is consistent with 
theoretical predictions (e.g. King \& Pringle 2006, 2007; King et al. 2008; 
Berti \& Volonteri 2008; Barausse 2012). 
The range of spin values obtained here and their evolution with 
redshift are remarkably similar to those indicated by the detailed numerical
study of Barausse (2012). The sources studied here are all associated with 
massive elliptical galaxies, and the evolution of the spins should 
be compared with those of the most massive galaxies at any redshift. 

As is clear from Figs. 4, 6 and 8 for radio galaxies 
and Figs. 5, 7, and 9 for radio loud quasars, 
and the results shown in columns 2 and 3 of Table 3, 
the new values obtained and
the range of values at a given redshift are very similar to those
obtained earlier using smaller samples of 19 FRII radio 
galaxies and 7 FRII radio loud quasars. Considering the combined
radio galaxy data set (of 52 new sources plus 19 previously studied sources)
and the combined quasar data set (of 23 new sources plus 7 previously 
studied sources),
the quasars spins appear to have a larger
range or scatter at a given redshift compared with radio galaxies 
(compare Figs. 4 \& 5; Figs. 6 \& 7; and Figs. 8 \& 9); this is least
pronounced for an Eddington magnetic field strength. 
The larger range of spin values of quasars at a given redshift compared
with radio galaxies is also reflected in the fits to the data. 
As can be seen from Table 3, the reduced 
$\chi^2$ of the quasar fit is about 25 \% larger than the radio galaxy fit for
an Eddington magnetic field strength and about 100 \% larger for the two 
other field
strengths considered,
which is caused by the larger scatter of quasar spin values at a 
given redshift; here 
the reduced $\chi^2$ is the value of $\chi^2/(N-2)$ since both the 
slope and y-intercept 
are obtained when fitting for spin as a function of redshift. 

The normalization and exponent of the empirically determined relation 
between beam power, L, and black hole mass, M, provides an indication of
whether the outflow is powered at least in part by spin energy extraction
from the black hole and, if so, whether the relevant magnetic field strength
depends upon the black hole mass. For FRII RG and FRII RLQ studied here, 
the normalization
of this relation indicates that the outflow is likely to be powered by black 
hole spin. 
The dependence of L on M indicates a field strength that is roughly independent
of black hole mass for RG and one that depends upon black hole mass for RLQ
with the dependence being roughly that expected for an Eddington field 
strength. 
For lower powered, primarily FRI sources associated with CD galaxies, the 
normalization of the L-M relation suggests that the outflow could be 
produced by the outflows
from the accretion disk though a contribution from spin energy extraction
is also quite possible. If spin energy extraction is involved, the field 
strength
is roughly independent of black hole mass. 

Considering black hole spin and black hole mass, there is no indication of
a strong dependence of spin on mass for the data as a whole or for subsets of 
the data. 

The spin values and range of values obtained here are consistent with those 
obtained with two independent methods: the use of the X-ray spectral properties 
of accretion disks associated with nearby AGN and the use of continuum-fitting
applied to stellar mass black holes. The results obtained here are in good
agreement with detailed numerical studies of supermassive black
hole spin evolution (e.g. Barausse 2012). And, there is now empirical
support for the fundamental relationship employed here to obtain
black hole spin values (Narayan \& McClintock 2012). Thus, the 
method of using outflows from AGN to determine supermassive black 
hole spin looks rather promsing.

\section*{Acknowledgments}
I would like to thank the referee for very helpful comments
and suggestions and Laura Brenneman, Kris Beckwith,  
Brian McNamara, Chris Reynolds, and Alexander Tchekhovskoy  
for interesting and helpful discussions. This work was
supported in part by Penn State University.

%\clearpage
%\end{document}

\begin{table*}
\begin{minipage}{140mm}
\caption{Black Hole Spins of FRII Radio Galaxies}   % title of Table
\label{tab:comp}        % is used to refer this table in the text
%\centering                          % used for centering table
\begin{tabular}{ccccccccc}   % centered columns (4 columns)
\hline\hline                    % inserts double horizontal lines
Source &   z & {$L_{44}$%
\footnote{The beam power is obtained from the 178 MHz radio power using 
the relationship given in the first line of Table 2 from 
Daly et al. (2012).}}  & {$M_8$% 
\footnote{Values are obtained from 
McLure et al. (2006).}}
& $j_M(B=B_{EDD})$ & $j_M(B=10^4G)$&$j_M(B \propto j)$
 \\
(1)    &    (2)         &    (3)          &   (4)           &  (5) 
      &       (6)    &(7)         \\
\hline                          % inserts single horizontal line
3C33	&	0.0595	&$	2.4	\pm	1.6	$&$	3.7	\pm	2.1	$&$	0.13	\pm	0.06	$&$	0.42	\pm	0.28	$&$	0.39	\pm	0.13	$ \\
3C192	&	0.0598	&$	1.1	\pm	0.9	$&$	2.6	\pm	1.5	$&$	0.11	\pm	0.05	$&$	0.40	\pm	0.28	$&$	0.38	\pm	0.13	$ \\
3C35	&	0.0677	&$	0.77	\pm	0.67	$&$	4.4	\pm	2.6	$&$	0.068	\pm	0.035	$&$	0.19	\pm	0.14	$&$	0.26	\pm	0.10	$ \\
3C285	&	0.0794	&$	1.1	\pm	0.9	$&$	3.4	\pm	2.0	$&$	0.093	\pm	0.046	$&$	0.30	\pm	0.21	$&$	0.33	\pm	0.12	$ \\
3C452	&	0.0811	&$	4.2	\pm	2.4	$&$	5.1	\pm	3.0	$&$	0.15	\pm	0.06	$&$	0.39	\pm	0.25	$&$	0.37	\pm	0.12	$ \\
3C326	&	0.0895	&$	2.2	\pm	1.5	$&$	2.4	\pm	1.4	$&$	0.16	\pm	0.07	$&$	0.60	\pm	0.40	$&$	0.46	\pm	0.16	$ \\
3C388	&	0.0908	&$	2.6	\pm	1.7	$&$	6.9	\pm	4.0	$&$	0.100	\pm	0.044	$&$	0.23	\pm	0.15	$&$	0.29	\pm	0.10	$ \\
3C321	&	0.096	&$	1.7	\pm	1.3	$&$	6.1	\pm	3.5	$&$	0.086	\pm	0.040	$&$	0.21	\pm	0.14	$&$	0.27	\pm	0.09	$ \\
3C236	&	0.0989	&$	1.9	\pm	1.4	$&$	6.5	\pm	3.8	$&$	0.088	\pm	0.041	$&$	0.21	\pm	0.14	$&$	0.27	\pm	0.09	$ \\
3C433	&	0.1016	&$	6.3	\pm	3.3	$&$	9.2	\pm	5.3	$&$	0.14	\pm	0.05	$&$	0.27	\pm	0.17	$&$	0.31	\pm	0.10	$ \\
3C223	&	0.1368	&$	3.5	\pm	2.2	$&$	3.0	\pm	1.7	$&$	0.18	\pm	0.07	$&$	0.61	\pm	0.40	$&$	0.47	\pm	0.15	$ \\
4C12.03	&	0.156	&$	3.3	\pm	2.0	$&$	7.1	\pm	4.2	$&$	0.11	\pm	0.05	$&$	0.25	\pm	0.16	$&$	0.30	\pm	0.10	$ \\
3C20	&	0.174	&$	13.0	\pm	5.2	$&$	3.8	\pm	2.2	$&$	0.30	\pm	0.11	$&$	0.93	\pm	0.58	$&$	0.58	\pm	0.18	$ \\
3C319	&	0.192	&$	6.8	\pm	3.5	$&$	2.1	\pm	1.2	$&$	0.29	\pm	0.11	$&$	1.2	\pm	0.8	$&$	0.66	\pm	0.21	$ \\
3C28	&	0.1952	&$	7.6	\pm	3.7	$&$	7.5	\pm	4.4	$&$	0.16	\pm	0.06	$&$	0.36	\pm	0.23	$&$	0.36	\pm	0.11	$ \\
3C349	&	0.205	&$	6.7	\pm	3.4	$&$	3.6	\pm	2.1	$&$	0.22	\pm	0.09	$&$	0.70	\pm	0.45	$&$	0.50	\pm	0.16	$ \\
3C132	&	0.214	&$	7.3	\pm	3.6	$&$	5.5	\pm	3.3	$&$	0.19	\pm	0.07	$&$	0.48	\pm	0.31	$&$	0.42	\pm	0.13	$ \\
3C436	&	0.2145	&$	9.4	\pm	4.3	$&$	7.1	\pm	4.2	$&$	0.19	\pm	0.07	$&$	0.42	\pm	0.27	$&$	0.39	\pm	0.12	$ \\
3C123	&	0.2177	&$	67	\pm	12	$&$	6.6	\pm	3.9	$&$	0.52	\pm	0.16	$&$	1.2	\pm	0.7	$&$	0.66	\pm	0.20	$ \\
3C171	&	0.2384	&$	12	\pm	5	$&$	3.9	\pm	2.4	$&$	0.29	\pm	0.10	$&$	0.87	\pm	0.55	$&$	0.56	\pm	0.18	$ \\
3C284	&	0.2394	&$	8.0	\pm	3.9	$&$	7.8	\pm	4.7	$&$	0.17	\pm	0.06	$&$	0.36	\pm	0.23	$&$	0.36	\pm	0.12	$ \\
3C79	&	0.2559	&$	21	\pm	7	$&$	6.0	\pm	3.6	$&$	0.30	\pm	0.10	$&$	0.73	\pm	0.46	$&$	0.51	\pm	0.16	$ \\
3C300	&	0.272	&$	14	\pm	6	$&$	3.6	\pm	2.2	$&$	0.33	\pm	0.12	$&$	1.0	\pm	0.7	$&$	0.61	\pm	0.19	$ \\
3C153	&	0.2769	&$	13	\pm	5	$&$	8.6	\pm	5.2	$&$	0.20	\pm	0.07	$&$	0.40	\pm	0.26	$&$	0.38	\pm	0.12	$ \\
3C438	&	0.29	&$	36	\pm	9	$&$	13	\pm	8	$&$	0.27	\pm	0.09	$&$	0.45	\pm	0.28	$&$	0.40	\pm	0.12	$ \\
3C299	&	0.367	&$	17	\pm	6	$&$	3.7	\pm	2.3	$&$	0.35	\pm	0.13	$&$	1.1	\pm	0.7	$&$	0.63	\pm	0.20	$ \\
4C14.27	&	0.392	&$	20	\pm	7	$&$	5.4	\pm	3.4	$&$	0.31	\pm	0.11	$&$	0.82	\pm	0.53	$&$	0.54	\pm	0.18	$ \\
3C42	&	0.395	&$	21	\pm	7	$&$	8.8	\pm	5.5	$&$	0.25	\pm	0.09	$&$	0.51	\pm	0.33	$&$	0.43	\pm	0.14	$ \\
3C16	&	0.405	&$	22	\pm	7	$&$	4.6	\pm	2.9	$&$	0.35	\pm	0.13	$&$	1.0	\pm	0.6	$&$	0.60	\pm	0.19	$ \\
3C274.1	&	0.422	&$	32	\pm	8	$&$	8.6	\pm	5.4	$&$	0.31	\pm	0.11	$&$	0.64	\pm	0.42	$&$	0.48	\pm	0.16	$ \\
3C457	&	0.428	&$	28	\pm	8	$&$	6.4	\pm	4.0	$&$	0.34	\pm	0.12	$&$	0.81	\pm	0.53	$&$	0.54	\pm	0.18	$ \\
3C46	&	0.4373	&$	25	\pm	8	$&$	15	\pm	10	$&$	0.21	\pm	0.07	$&$	0.31	\pm	0.21	$&$	0.34	\pm	0.11	$ \\
3C341	&	0.448	&$	25	\pm	8	$&$	10	\pm	7	$&$	0.25	\pm	0.09	$&$	0.47	\pm	0.31	$&$	0.41	\pm	0.13	$ \\
3C200	&	0.458	&$	27	\pm	8	$&$	8.8	\pm	5.6	$&$	0.29	\pm	0.10	$&$	0.58	\pm	0.38	$&$	0.46	\pm	0.15	$ \\
3C295	&	0.4614	&$	130	\pm	19	$&$	29	\pm	18	$&$	0.35	\pm	0.12	$&$	0.40	\pm	0.26	$&$	0.38	\pm	0.12	$ \\
3C19	&	0.482	&$	29	\pm	8	$&$	14	\pm	9	$&$	0.23	\pm	0.08	$&$	0.37	\pm	0.24	$&$	0.36	\pm	0.12	$ \\
3C225B	&	0.58	&$	76	\pm	12	$&$	10	\pm	7	$&$	0.45	\pm	0.15	$&$	0.85	\pm	0.58	$&$	0.55	\pm	0.19	$ \\
3C49	&	0.6207	&$	42	\pm	10	$&$	13	\pm	9	$&$	0.29	\pm	0.11	$&$	0.49	\pm	0.34	$&$	0.42	\pm	0.15	$ \\
3C277.2	&	0.766	&$	85	\pm	13	$&$	11	\pm	8	$&$	0.45	\pm	0.17	$&$	0.83	\pm	0.60	$&$	0.54	\pm	0.20	$ \\
3C340	&	0.7754	&$	65	\pm	11	$&$	11	\pm	8	$&$	0.40	\pm	0.15	$&$	0.72	\pm	0.52	$&$	0.51	\pm	0.18	$ \\
3C352	&	0.806	&$	83	\pm	13	$&$	16	\pm	11	$&$	0.37	\pm	0.14	$&$	0.57	\pm	0.42	$&$	0.45	\pm	0.17	$ \\
3C263.1	&	0.824	&$	130	\pm	18	$&$	19	\pm	14	$&$	0.43	\pm	0.16	$&$	0.60	\pm	0.44	$&$	0.46	\pm	0.17	$ \\
3C217	&	0.8975	&$	98	\pm	14	$&$	6.6	\pm	4.9	$&$	0.63	\pm	0.24	$&$	1.5	\pm	1.1	$&$	0.73	\pm	0.28	$ \\
3C175.1	&	0.92	&$	110	\pm	15	$&$	12	\pm	9	$&$	0.49	\pm	0.19	$&$	0.85	\pm	0.64	$&$	0.55	\pm	0.21	$ \\
3C252	&	1.105	&$	170	\pm	25	$&$	20	\pm	16	$&$	0.47	\pm	0.19	$&$	0.63	\pm	0.51	$&$	0.48	\pm	0.19	$ \\
3C368	&	1.132	&$	240	\pm	44	$&$	28	\pm	23	$&$	0.48	\pm	0.20	$&$	0.54	\pm	0.44	$&$	0.44	\pm	0.18	$ \\
3C266	&	1.272	&$	220	\pm	38	$&$	23	\pm	20	$&$	0.50	\pm	0.22	$&$	0.62	\pm	0.53	$&$	0.47	\pm	0.20	$ \\
3C13	&	1.351	&$	260	\pm	47	$&$	40	\pm	34	$&$	0.41	\pm	0.18	$&$	0.39	\pm	0.34	$&$	0.38	\pm	0.16	$ \\
4C13.66	&	1.45	&$	260	\pm	47	$&$	16	\pm	14	$&$	0.66	\pm	0.30	$&$	0.99	\pm	0.89	$&$	0.60	\pm	0.27	$ \\
3C241	&	1.617	&$	370	\pm	85	$&$	37	\pm	35	$&$	0.51	\pm	0.24	$&$	0.50	\pm	0.47	$&$	0.43	\pm	0.20	$ \\
3C470	&	1.653	&$	290	\pm	59	$&$	28	\pm	26	$&$	0.53	\pm	0.25	$&$	0.60	\pm	0.57	$&$	0.47	\pm	0.22	$ \\
3C294	&	1.786	&$	440	\pm	110	$&$	29	\pm	28	$&$	0.64	\pm	0.32	$&$	0.71	\pm	0.70	$&$	0.51	\pm	0.25	$ \\
\hline                      
\end{tabular}
\end{minipage}
\end{table*}

\begin{table*}
\begin{minipage}{140mm}
\caption{Black Hole Spins of FRII Radio Loud Quasars}   % title of Table
\label{tab:comp}        % is used to refer this table in the text
%\centering                          % used for centering table
\begin{tabular}{ccccccccc}   % centered columns (4 columns)
\hline\hline    
Source &   z & {$L_{44}$%
\footnote{The beam power is obtained from the 178 MHz radio power using 
the relationship given in the first line of Table 2 from 
Daly et al. (2012).}}  & {$M_8$% 
\footnote{Values are obtained from 
McLure et al. (2006).}}
& $j_M(B=B_{EDD})$ & $j_M(B=10^4G)$&$j_M(B \propto j)$
 \\
(1)    &    (2)         &    (3)          &   (4)           &  (5) 
      &       (6)    &(7)         \\
\hline                          % inserts single horizontal line
3C109	&	0.3056	&$	21	\pm	7	$&$	2.0	$&$	0.53	\pm	0.26	$&$	2.3	\pm	2.1	$&$	0.90	\pm	0.42	$ \\
3C249.1	&	0.311	&$	12	\pm	5	$&$	20	$&$	0.13	\pm	0.06	$&$	0.17	\pm	0.16	$&$	0.25	\pm	0.12	$ \\
3C351	&	0.371	&$	20	\pm	7	$&$	32	$&$	0.13	\pm	0.06	$&$	0.14	\pm	0.13	$&$	0.22	\pm	0.10	$ \\
3C215	&	0.411	&$	23	\pm	7	$&$	2.0	$&$	0.56	\pm	0.27	$&$	2.4	\pm	2.2	$&$	0.92	\pm	0.43	$ \\
3C47	&	0.425	&$	49	\pm	10	$&$	16	$&$	0.29	\pm	0.14	$&$	0.43	\pm	0.40	$&$	0.39	\pm	0.18	$ \\
3C263	&	0.652	&$	68	\pm	12	$&$	13	$&$	0.38	\pm	0.18	$&$	0.64	\pm	0.59	$&$	0.48	\pm	0.22	$ \\
3C207	&	0.684	&$	71	\pm	12	$&$	3.2	$&$	0.77	\pm	0.36	$&$	2.6	\pm	2.4	$&$	0.97	\pm	0.45	$ \\
3C196	&	0.871	&$	410	\pm	100	$&$	40	$&$	0.52	\pm	0.25	$&$	0.50	\pm	0.46	$&$	0.42	\pm	0.20	$ \\
3C309.1	&	0.904	&$	160	\pm	22	$&$	13	$&$	0.57	\pm	0.27	$&$	0.97	\pm	0.89	$&$	0.59	\pm	0.27	$ \\
3C336	&	0.927	&$	100	\pm	15	$&$	16	$&$	0.41	\pm	0.19	$&$	0.63	\pm	0.58	$&$	0.47	\pm	0.22	$ \\
3C245	&	1.029	&$	160	\pm	23	$&$	25	$&$	0.41	\pm	0.19	$&$	0.49	\pm	0.45	$&$	0.42	\pm	0.19	$ \\
3C212	&	1.049	&$	190	\pm	28	$&$	16	$&$	0.56	\pm	0.26	$&$	0.84	\pm	0.77	$&$	0.55	\pm	0.25	$ \\
3C186	&	1.063	&$	210	\pm	33	$&$	32	$&$	0.41	\pm	0.19	$&$	0.44	\pm	0.41	$&$	0.40	\pm	0.18	$ \\
3C208	&	1.109	&$	230	\pm	40	$&$	25	$&$	0.49	\pm	0.23	$&$	0.59	\pm	0.55	$&$	0.46	\pm	0.21	$ \\
3C204	&	1.112	&$	170	\pm	25	$&$	32	$&$	0.38	\pm	0.18	$&$	0.40	\pm	0.37	$&$	0.38	\pm	0.18	$ \\
3C190	&	1.197	&$	240	\pm	43	$&$	5.0	$&$	1.1	\pm	0.5	$&$	3.0	\pm	2.8	$&$	1.0	\pm	0.5	$ \\
4C16.49	&	1.296	&$	220	\pm	37	$&$	63	$&$	0.30	\pm	0.14	$&$	0.23	\pm	0.21	$&$	0.29	\pm	0.13	$ \\
3C181	&	1.382	&$	330	\pm	71	$&$	40	$&$	0.47	\pm	0.22	$&$	0.44	\pm	0.41	$&$	0.40	\pm	0.19	$ \\
3C14	&	1.469	&$	240	\pm	44	$&$	25	$&$	0.51	\pm	0.24	$&$	0.61	\pm	0.56	$&$	0.47	\pm	0.22	$ \\
3C205	&	1.534	&$	330	\pm	71	$&$	40	$&$	0.47	\pm	0.22	$&$	0.45	\pm	0.41	$&$	0.40	\pm	0.19	$ \\
3C432	&	1.805	&$	440	\pm	110	$&$	130	$&$	0.31	\pm	0.15	$&$	0.16	\pm	0.15	$&$	0.24	\pm	0.11	$ \\
3C191	&	1.9523	&$	600	\pm	180	$&$	50	$&$	0.56	\pm	0.27	$&$	0.48	\pm	0.44	$&$	0.41	\pm	0.19	$ \\
3C9	&	2.012	&$	940	\pm	350	$&$	63	$&$	0.63	\pm	0.31	$&$	0.47	\pm	0.44	$&$	0.41	\pm	0.19	$ \\
\hline			
\end{tabular}
\end{minipage}
\end{table*}

\begin{table*}
\begin{minipage}{140mm}
\caption{Comparison of Slopes for Log$(j)$ as a function of Log$(1+z)$ for FRII Radio Galaxies and Quasars}   % title of Table
\label{tab:comp}        % is used to refer this table in the text
%\centering                          % used for centering table
\begin{tabular}{lcccccc}   % centered columns (4 columns)
\hline\hline  
	&	FRII RG Slope & $\chi^2$ of Fit	&	FRII RLQ 
Slope 	& $\chi^2$ of Fit &	FRII (RG + RLQ) Slope & $\chi^2$ of Fit	\\

\hline                          % inserts single horizontal line
Number of Sources&71&&30&&101\\
$B=10^4 G$	&$0.60 \pm 0.22$ & 33.3 &$-0.23 \pm 0.71$& 27.3 &$0.44 \pm 0.20$& 61.8 \\
$B=B_{EDD}$	&$1.37 \pm 0.15$&54.1&$0.96 \pm 0.36$ & 24.5& 
$1.34 \pm 0.13$& 80.0 \\
$B \propto j$	&$0.30 \pm 0.11$&33.5 & $-0.12 \pm 0.36$&27.3& 
$0.22 \pm 0.10$& 62.1 \\
\hline			
\end{tabular}
\end{minipage}
\end{table*}

\begin{table*}
\begin{minipage}{140mm}
\caption{Comparison of Slopes for Log$(j)$ as a function of Log$(1+z)$ for FRII Sources}   % title of Table
\label{tab:comp}        % is used to refer this table in the text
%\centering                          % used for centering table
\begin{tabular}{lccccc}   % centered columns (4 columns)
\hline\hline  
	&	Updated Slopes	&	Prior Slopes	&	Slope for Fixed Mass	&	Slope for Fixed Mass	&	Slope for Fixed Mass	\\
	&	71 RG + 30 RLQ	&	19 RG + 7 RLQ	&	
Constant Log {$M_1$%
\footnote{The weighted mean of the Log of the black hole mass for the 55 sources studied previously is $1.16 \pm 0.03$, where the mass is in units of
$10^8 M_{\odot}$. Only FRII sources with Log(M) within one 
one sigma of this value were included; this left 2 radio loud quasars 
and 17 radio galaxies.}}	&	Constant {$M_2$%
\footnote{The weighted mean of the black hole masses of the 
55 sources studied previously (including 19 FRII radio galaxies, 7 FRII radio loud
quasars, and 29 radio sources associated with CD galaxies)
is $(6.5 \pm 0.6) \times 10^8 M_{\odot}$. Only FRII sources with 
masses within one sigma of this value were included; this left 6 radio 
loud quasars and 17 radio galaxies. }} 	&	Constant {$M_3$%
\footnote{The unweighted mean of the black hole masses of the 55 sources
studied previously is $24.4 \times 10^8 M_{\odot}$. Only FRII sources with
masses within one sigma of this value were included; this left 5 radio loud
quasars and 13 radio galaxies. }}	\\

	&	N = 101	&	N = 26	&	N = 19	&	N = 23	&	N = 18	\\

(1)    &    (2)         &    (3)          &   (4)           &  (5)  & (6)   \\
\hline                          % inserts single horizontal line

$B=10^4 G$	&$	0.44 \pm 0.20	$&$	0.86 \pm 0.36	$&$	0.84 \pm 0.60	$&$	0.71 \pm 0.60	$&$	0.70 \pm 0.26	$\\
$B=B_{EDD}$	&$	1.3 \pm 0.1	$&$	1.1 \pm 0.2	$&$	1.7 \pm 0.4	$&$	1.6 \pm 0.4	$&$	0.94 \pm 0.19	$\\
$B \propto j$	&$	0.22 \pm 0.10	$&$	0.43 \pm 0.18	$&$	0.42 \pm 0.30	$&$	0.35 \pm 0.30	$&$	0.35 \pm 0.13	$\\
\hline			
\end{tabular}
\end{minipage}
\end{table*}

\begin{table*}
\begin{minipage}{140mm}
\caption{Unweighted Fits to ${\rm Log}(j) = m {\rm Log}(M_8)~+b$}   
% title of Table
\label{tab:comp1}        % is used to refer this table in the text
%\centering                          % used for centering table
\begin{tabular}{lrlcrlcrl}   % centered columns (4 columns)
\hline\hline  
	&	71 &FRII RG &&30 &FRII RLQ&	& 29 &CD \\
\hline 
&m~~~~~~~~&~~~~~~~~~b&&m~~~~~~~~&~~~~~~~~b&&m~~~~~~~~&~~~~~~~~~b\\
                       % inserts single horizontal line
$B=10^4 G$	&$-0.10 \pm 0.07$& $-0.18 \pm 0.08$&& $-0.66 \pm 0.08$&
$0.62 \pm 0.12$&&$0.06 \pm 0.23$& $-1.20 \pm 0.29$ \\
$B=B_{EDD}$	&$0.40 \pm 0.07$&$ -0.96 \pm 0.08$&&$-0.16 \pm 0.08$& 
$-0.16 \pm 0.12$&&$ 0.56 \pm 0.23$&$-1.97 \pm 0.29$ \\
$B \propto j$	& $-0.05 \pm 0.04$& $-0.31 \pm 0.04$&&$-0.33 \pm 0.04$&
$0.09 \pm 0.06$&&$0.03 \pm 0.11$&$-0.83 \pm 0.14$\\
\hline			
\end{tabular}
\end{minipage}
\end{table*}

\end{document}